\RequirePackage{fix-cm}
\documentclass[smallextended]{svjour3}       
\smartqed  
\usepackage{graphicx}

\usepackage{amsmath}
\usepackage{amssymb}

\usepackage{fancyvrb}
\usepackage{hyperref}
\usepackage[numbers]{natbib}

\begin{document}

\title{How well do Network Models predict Observations? On the Importance of Predictability in Network Models}
	

\titlerunning{Predictability in Network Models}        

\author{Jonas Haslbeck  \and Lourens Waldorp}


\institute{Jonas Haslbeck \at
              University of Amsterdam\\
              Psychological Methods\\
              Nieuwe Achtergracht 129-B\\
              1018 WT, Amsterdam\\
              \email{jonashaslbeck@gmal.com}           
       }

\date{Received: date / Accepted: date}

\maketitle

\begin{abstract}

Network models are an increasingly popular way to abstract complex psychological phenomena. While the study of the \textit{structure} of network models has led to many important insights, little attention is paid to how well they \textit{predict} observations. This is despite the fact that predictability is crucial for judging the \textit{practical relevance} of edges: for instance in clinical practice, predictability of a symptom indicates whether a an intervention on that symptom through the symptom network is promising. We close this methodological gap by introducing nodewise predictability, which quantifies how well a given node can be predicted by all other nodes it is connected to in the network. In addition, we provide fully reproducible code examples of how to compute and visualize nodewise predictability both for cross-sectional and time-series data.

\end{abstract}

\section{Introduction}\label{intro}

\begin{figure*}
	\centering
	\includegraphics[width=0.75\textwidth]{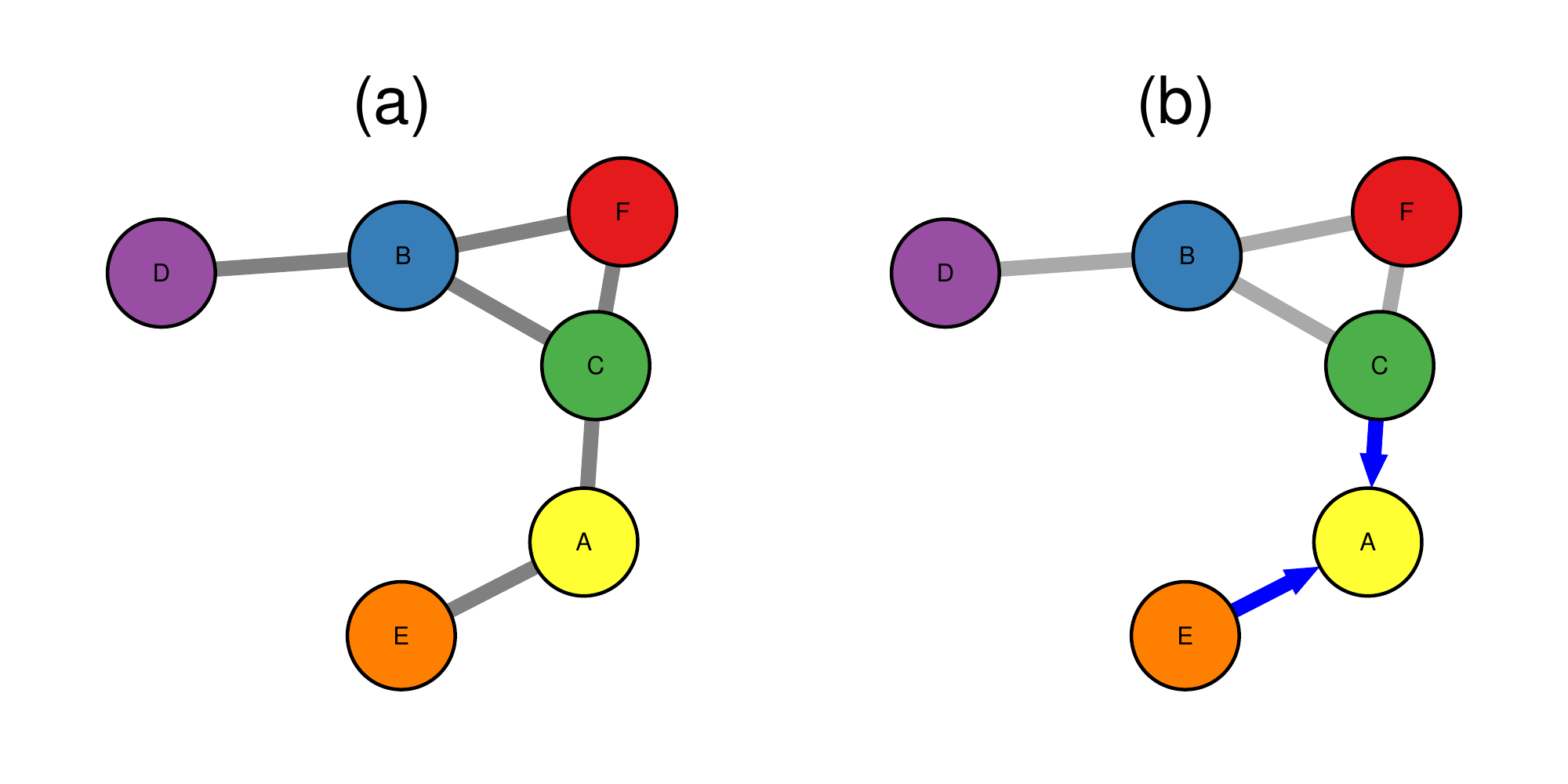}
	\caption{(a) Example network with six nodes. An edge between two nodes indicates a pairwise interaction between those two nodes (b) Illustration of predicting node A by all its neighboring nodes (C and E).}
	\label{figure_1}       
\end{figure*}

Network models graphically describe interactions between a potentially large number variables: each variable is represented as a dot (node) and interactions are represented by lines (edges) connecting the nodes (for an illustration see Figure \ref{figure_1} a). These models have been a popular way to abstract complex systems in a large variety of disciplines such as statistical mechanics \cite{Albert_statistical_2002}, biology \cite{friedman_using_2000}, neuroscience \cite{huang_learning_2010} and are recently also applied in psychology \cite{costantini2015state} and psychiatry \cite{Borsboom_Network_2013}. 

Particularly in psychology, network models are attractive because many psychological phenomena are considered to depend on a large number of variables and interactions between them. In this situation, the graphical representation ensures that the model can be understood intuitively even if the number of variables is large.  In addition, network models open up the possibility to study the network structure: for instance, one can use network summary measures like density or centrality to describe the global structure of the network \cite{newman2010networks}. These could allow inferences about the behavior of the whole network that would not be possible from the edge parameters alone. One could also run generative models on the network, e.g. diffusion models of diseases to explain how symptoms of psychological disorders activate each other \cite{shulgin1998pulse}. 

Currently, most applications are in the field of clinical psychology \cite[e.g.][]{fried2016good, fried2015loss, beard2016network, mcnally2015mental, boschloo2015network} but network models are also applied in other subfields such as health psychology \cite{kossakowski2016application} and personality psychology \cite{cramer2012dimensions, costantini2015state}. While initially they were used to model cross-sectional data, there is increasing interest in analyzing data obtained using the Experience Sampling Method (ESM), which consists of repeated measurements of the same person \cite[e.g.,][]{bringmann2013network, pe2015emotion}. The focus in these papers is the global network structure and the connectedness of specific nodes in the network, which provide a new perspective on many psychological phenomena. For instance, Cramer and colleagues \cite{cramer2010comorbidity} suggested an alternative view on the concept of comorbidity by analyzing how symptoms of different psychological disorders are connected to each other.

The key idea of this paper is to analyze the \textit{predictability} of nodes in the network \textit{in addition} to the network structure. By predictability of node A we mean how well node A can be predicted by all its neighboring nodes in the network (see Figure 1b). The predictability of nodes is important for several reasons:

\begin{enumerate}
	\item The edges connected to node A should be interpreted taking into account how much of the variance of A is explained by the edges connected to A. For instance, edges will be interpreted differently, depending on whether 0.5\% or 50\% of the variance of A is explained. This issue is particularly important for networks estimated on a large number of observations, where small edge weights can be detected that are practically meaningless.
	\item In many areas of psychology the goal is to design effective interventions. Using the predictability measure of node A, one can estimate how much we can influence this node by intervening on nodes that are connected to it.
	\item Predictability across nodes tells us whether a (part of a) network is largely determined by itself through strong mutual interactions between nodes (high predictability) or whether it is mostly determined by other factors that are not included in the network (low predictability).
\end{enumerate}

The problem addressed here is similar to the problem of modeling only the covariance matrix in Structural Equation Modeling (SEM) \cite{byrne2013structural}: one might find a model that perfectly fits the covariance matrix, but if the variance of variables is much larger than their covariance, the model might be meaningless in practice. 

Predictability in general cannot be inferred by the network structure but has to be computed from the network model and the data. Unfortunately, currently there is no easy-to-use tool available for researchers to compute and present predictability in network models. In the present paper, we close this methodological gap by making the following contributions:

\begin{enumerate}
	\item We present a method to compute easy to interpret nodewise predictability measures for state-of-the-art network models (Section \ref{section_methods}).
	\item We provide a step-by-step description of how to use the R-packages \textit{mgm} and \textit{qgraph} to compute and visualize nodewise predictability, both for cross-sectional (Section \ref{section_crossSec}) and time-series networks (Section \ref{section_temporal}). The provided code is fully reproducible, which means that the reader can run the code and reproduce all figures while reading. The data in our applications are from two published studies and will be downloaded automatically with the provided code.
	\end{enumerate}

\section{Methods}\label{section_methods}

In order to determine the predictability of a given node A, we need to know which nodes are connected to A in the network model. Therefore the first step is to estimate a network model, which we describe in Section \ref{section_model}. In a second step, we use the network model to predict the given node A by the nodes that are connected to it (its neighbors). In Section \ref{section_prediction}, we describe in detail how to compute these predictions. Finally, we quantify how close these predictions are to the actual values of A. The closer the predictions are to the actual values, the higher the predictability of A. A description of predictability measures for both continuous and categorical variables is given in Section \ref{section_measures}. In Section \ref{section_pred_par} we discuss the relationship between the predictability and the parameters of the network model. Finally we describe the data (\ref{sectio_datainfo}) that is used in the application examples in Sections \ref{section_crossSec} and \ref{section_temporal}.

\subsection{Network Models}\label{section_model}

We model cross-sectional data using pairwise Mixed Graphical Models (MGMs) \cite{yang_mixed_2014, haslbeck2015structure}, which generalize well-known exponential family distributions such as the multivariate Gaussian distribution or the Ising model \cite{wainwright_graphical_2008}. This is the model used in all papers mentioned in the introduction. MGMs are estimated via $\ell_1$-regularized (LASSO) neighborhood regression as implemented in the R-package mgm by the authors \cite{haslbeck2015mgm}. In this approach, one estimates the neighborhood of each node and combines all neighborhoods to obtain the complete graph (network) \cite{meinshausen_high-dimensional_2006}. The \textit{neighborhood} of a node is the set of nodes that is connected to that node. For example, in Figure 1(a), the neighborhood of node A consists of the nodes E and C. The $\ell_1$ regularization ensures that spurious edge-parameters are put to exactly zero, which makes the network model easier to interpret. The parameter that controls the strength of the regularization is selected via 10 fold cross validation.

For time-series data we use the Vector Autoregressive (VAR) model, which is a popular model for multivariate time  series in many disciplines \cite[see e.g.][]{hamilton_time_1994, pfaff_analysis_2008}. The VAR model is different from the MGM in that associations are now defined between time-lagged variables. Specifically, in its simplest form with a time-lag of order one, in this model \text{all} variables $X^{t-1}$ at time $t - 1$ are regressed on \textit{each} of the variables $X^{t}_i$ at time $t$, where $i$ indexes different variables. Note that this also includes the variable $X_s$ itself at an earlier time point: that is, one predicts $X_s^t$ at time $t$ by itself and all other variables at time $t - 1$. For the analyses in this paper we use the implementation of mixed VAR models in the R-package mgm \cite{haslbeck2015mgm}.

\subsection{Making Predictions}\label{section_prediction}

We are interested in how well a node can be predicted by all adjacent nodes in the network. This means that we would like to compute the mean of the conditional distribution of the node at hand given all its neighbors. To provide understanding of what this means exactly, we show how to compute predictability for the node $A$ in Figure \ref{figure_1} (b), for (1) the case of $A$ being a continuous-Gaussian variable and (2) the case of $A$ being binary.  

We begin with (1): the conditional mean of $A$ given its neighbors $C$ and $E$, which is given by

\begin{equation}\label{gauss_model}
P(A = x |C, E)  = \frac{1}{\sqrt{2\pi }\sigma } \exp \left  \{  - \frac{(x - \mu)^2}{2 \sigma^2}   \right  \},
\end{equation}

\noindent
where the mean $\mu = \beta_0 + \beta_C C + \beta_E E$ is a linear combination of the two neighbors $C$ and $E$. This conditional distribution is obtained from the multivariate exponential family distribution of the MGM, for details see \cite{yang_mixed_2014, haslbeck2015structure}. This prediction problem corresponds to the familiar linear regression problem with Gaussian noise. Now, how to make predictions? Let's say the intercept is $\beta_0 = 0.25$ and $\beta_C = 0.1, \beta_E = -0.5$. Then if the $i^{th}$ case in the sample is $C_i = 2, E_i=1$, then for the  $i^{th}$ sample of $A$ we predict $A_i = 0.25 + 0.1 \times 2 - 0.5 \times 1 = -0.05$. A measure of predictability should evaluate how close this is the actual observation for node $A_i$.

In example (2), where $A$ is categorical, we compute a predicted probability for each category using a multinomial distribution

\begin{equation}\label{multinom_model}
P(A = k|C, E)  = \frac{\exp\{ \mu_k  \} }{\sum_{l=1}^K \exp\{ \mu_k \}},
\end{equation}

\noindent
where $k$ indicates the category, $K$ is the number of categories and $\mu_k = \beta_{0k} + \beta_{Ck} C +  \beta_{Ek} E$. Now let's assume $A$ is binary ($K = 2$) and we have $\beta_{01} = 0, \beta_{C1} = 0.5, \beta_{E1} = 1$ and  $\beta_{02} = 0, \beta_{C2} = -0.5, \beta_{E2} = -1$ and if for the $i^{th}$ cases we have $C_i = 1$ and $E_i = 1$. When filling in the numbers in equation (\ref{multinom_model}) we get $P(A = 1|C, E) \approx 0.95$ and $P(A = 2|C, E) \approx 0.05$, and predict category $k=1$ for the $i^{th}$ sample of $A$, because $0.95 > \frac{1}{2}$. Of course, all probabilities have to add up to 1, so we have $1 - P(A = 1|C, E) = P(F = 2|C, E)$. This direct approach of modeling the probabilities of categories is possible due to the regularization used in estimation \cite[see e.g.,][]{hastie2015statistical}, otherwise this model would not be identified. Note that predicting $A$ by all its neighbors is the same as predicting A by all nodes in the network. This is because all nodes that are \textit{not} in the neighborhood of $A$ have a zero weight associated to them in the regression equation on $A$ (\ref{gauss_model} or \ref{multinom_model}) and can hence be dropped.

In the case of other exponential family distributions, such as Poisson or Exponential, one similarly uses the univariate conditional distribution to make predictions \cite{yang_mixed_2014}. Importantly, the joint distribution of the MGM can be represented as a factorization of $p$ conditional distributions and hence our method to compute predictions is based on a proper representation of the joint distribution. Indeed, this factorization is used when estimating the MGM in the neighborhood regression approach (see Section \ref{section_model}).

\subsection{Quantifying Predictability}\label{section_measures}

After computing predictions, we would like to know how close these are to the observed values in the data. Because it is of interest how well a given node can be predicted \textit{by all other nodes in the network}, we need to remove any effects of the intercept (continuous variables) and the marginal (categorical variables). The marginal indicates the probabilities of categories, when ignoring all other variables. For example, the marginal of a binary variable is described by relative frequency of observing category 1,  e.g. $P(X = 1) = 0.7$. 

\subsubsection{Predictability in Continuous Variables}

For continuous data, we choose the proportion of explained variance as predictability measure as it is well-known in the literature and easy to interpret:

$$ R^2_A = 1 - \frac{var(\hat{A} - A)  }{var(A)}, $$

\noindent
where $var$ is the variance, $\hat{A}$ is a vector of predictions for $A$ as described in Section \ref{section_prediction}, and $A$ is the vector of observed values in the data. In order to remove any influences of the intercepts, all variables are centered to mean zero. Hence, all intercepts will be zero and cannot affect to the predictability measure. Thus, we can interpret $R^2$ as follows: a value of 0 means that a node cannot be predicted at all by its neighboring nodes in the network, whereas a value of 1 means that a node can be perfectly predicted by its neighboring nodes.

\subsubsection{Predictability in Categorical Variables}

For categorical variables it is slightly more difficult to get a measure with the same interpretation as the $R^2$ for continuous variables, because there is no way to center categorical variables. The following example shows that it is, however, important to somehow take the marginal into account: let's say we have 100 observations of a binary variable $A$ and observe 10 1s and 90 1s. This means that the marginal probabilities of $A$ are $p_0 = 0.1$ and $p_1 = 0.9$. Now, if all other nodes contribute nothing to predicting whether there is a 0 or 1 present in case $A_i$, one can just predict a 1 for all cases and get a proportion of correct classification (or accuracy, see below) of $90 \% $. For our purpose of determining how well a node can be predicted by all other nodes, this is clearly misleading, because actually \textit{nothing} is predicted by all other nodes. We therefore compute a \textit{normalized accuracy} that removes the accuracy that is achieved by the trivial prediction using marginal of the variable ($p_1 = 0.9$) alone:

Let $\mathcal{A} = \frac{1}{n} \sum_{i=1}^{n} \mathbb{I} (y_i = \hat{y}_i)$ be the proportion of correct predictions, the accuracy, and let $p_0, p_1, \dots p_m$ be the marginal probabilities of the categories, where $\mathbb{I}$ is the indicator function for the event   $F_i = \hat{F}_i$. In the binary case these are $p_0$ and $p_1 = 1 - p_0$. We then define normalized accuracy as

$$ \mathcal{A} _{norm}  = \frac{\mathcal{A}  - \max\{p_0, p_1, \dots, p_m\}}{1 - \max\{p_0, p_1, \dots, p_m\}}. $$

Hence, $ \mathcal{A} _{norm}$ indicates how much the node at hand can be predicted by all other nodes in the network, \textit{beyond} what is trivially predicted by the marginal distribution. $ \mathcal{A} _{norm} = 0$ means that none of the other nodes adds anything to the marginal in predicting the node at hand, while $ \mathcal{A} _{norm} = 1$ means that all other nodes perfectly predict the node at hand (together with the marginal). 

Let's return to the above example: in contrast to the high accuracy of $\mathcal{A}  = 0.9$, the normalized accuracy $\mathcal{A} _{norm}$ is zero, indicating that the node at hand cannot be predicted by other nodes in the network. However, notice that both $\mathcal{A} $ and $\mathcal{A} _{norm}$ are important for interpretation. For instance if we have a marginal of $p_1 = .9$ in a binary variable, then it is less impressive if all other predictors account for $80\%$ of the remaining accuracy that can be achieved ($.98$ instead of $.9$) than in a situation where $p_1 = .5$, where accounting $80\%$ of the remaining accuracy would mean an improvement from $.5$ to $.9$. We therefore visualize both $\mathcal{A}$ and $\mathcal{A} _{norm}$ for the binary variable in Figure \ref{figure_2}.

 \subsection{Predictability and Model Parameters}\label{section_pred_par}
 
Given the above definition of measures of predictability, it is evident that there is a close relationship between the parameters of the network model and predictability: if a node is not connected to any other node then the explained variance/normalized accuracy of this node \textit{has} to be 0. Also, the more edges are connected to a node, the higher predictability tends to be. There is a strong linear relationship between predictability and edge parameters for Gaussian Graphical Models (GGM), where the edge parameters (partial correlation) are restricted to $[-1,1]$. This linear relationship is much weaker for models including categorical variables, where the model parameters are only constrained to be finite. 

This implies that also centrality measures (like degree centrality), which are a function of edge parameters, are strongly correlated with predictability for GGMs, but much less for MGMs \cite[e.g., ][]{haslbeck_Fried_2016}. However, note that even if a given centrality measure would correlate perfectly with predictability, it would not be a substitute, because it would only allow us to order nodes by predictability but would \textit{not} tell us the predictability of any node. Hence, while centrality measures are related to predictability, they are not a good proxy for predictability.

 \subsection{Application to Datasets}\label{sectio_datainfo}
 
We illustrate how to compute and visualize nodewise predictability for network models for both cross-sectional and time-series data. We use a cross-sectional dataset from \cite{fried2015loss} (N = 515) with 11 variables on the relationship on bereavement and depressive symptoms. In order to illustrate predictability for the VAR model, we use a dataset consisting of up to 10 daily measurements of nine variables related to mood over a long period of time (N = 1478) of a single individual \cite{wichers2016critical}. A detailed description of the time-series data can be found in \cite{kossakowski2016data}.

\section{Predictability in Cross-Sectional Networks}\label{section_crossSec}

Here we show how to obtain the proposed predictability measures using the mgm package. We will give the code below so all steps can be reproduced exactly by the reader.

First, we download the preprocessed data. The raw data and the preprocessing file can be found in the same Github repository.

\begin{Verbatim}[commandchars=\\\{\}, fontsize=\small]
library(httr)
url <- "https://github.com/jmbh/NetworkPrediction/raw/master/Fried2015_nD.RDS"
GET(url, write_disk("Fried2015.RDS", overwrite=TRUE))
datalist <- readRDS("Fried2015.RDS")
\end{Verbatim}

Next, we fit a MGM using the mgm-package:

\begin{Verbatim}[commandchars=\\\{\}, fontsize=\small]
library(mgm)
fit_obj <- mgm(data = datalist$data, 
	       type = c(rep("g", 11), "c"), 
	       lev = c(rep(1, 11), 2), 
	       ruleReg = "OR",
	       k = 2, binarySign = TRUE)
\end{Verbatim}

In addition to the data, one has to specify the type and the number of categories for each variable. The remaining arguments are tuning parameters and are selected such that the original results in \cite{fried2015loss} are reproduced. For the general usage of the mgm package see \cite{haslbeck2015mgm}. After estimating the model, which is saved in \verb|fit_obj|, we use the \verb|predict()| function to compute the predictability for each node in the network. For categorical variables, we specify the predictability measures accuracy / correct classification (\verb|"CC"|) and normalized accuracy (\verb|"nCC"|). In addition, we request the accuracy of the intercept (marginal) model (\verb|"CCmarg"|), which we will use to visualize the accuracy decomposition in intercept model and the contribution of other variables. For continuous variables, we specify explained variance (\verb|"R2"|)  as predictability measure.

\begin{Verbatim}[commandchars=\\\{\}, fontsize=\small]
pred_obj <- predict(fit_obj, datalist$data,
		    errorCat = c("CC", "nCC", "CCmarg"),
		    errorCon = c("R2"))
\end{Verbatim}

To display both the accuracy of the intercept model and the normalized accuracy (contribution by other variables), we require a list for the ring-segments and a list for the corresponding colors:

\begin{Verbatim}[commandchars=\\\{\}, fontsize=\small]
error_list <- list() # List for ring-segments
for(i in 1:11)  error_list[[i]] <- pred_obj$errors[i, 2]
error_list[[12]] <- c(p_obj$errors[12,5], p_obj$errors[12,3]-p_obj$errors[12,5])

color_list <- list() # List for Colors
for(i in 1:11) color_list[[i]] <- "#90B4D4"
color_list[[12]] <- c("#ffa500", "#ff4300")

\end{Verbatim}

\begin{figure*}
	\centering
	\includegraphics[width=0.75\textwidth]{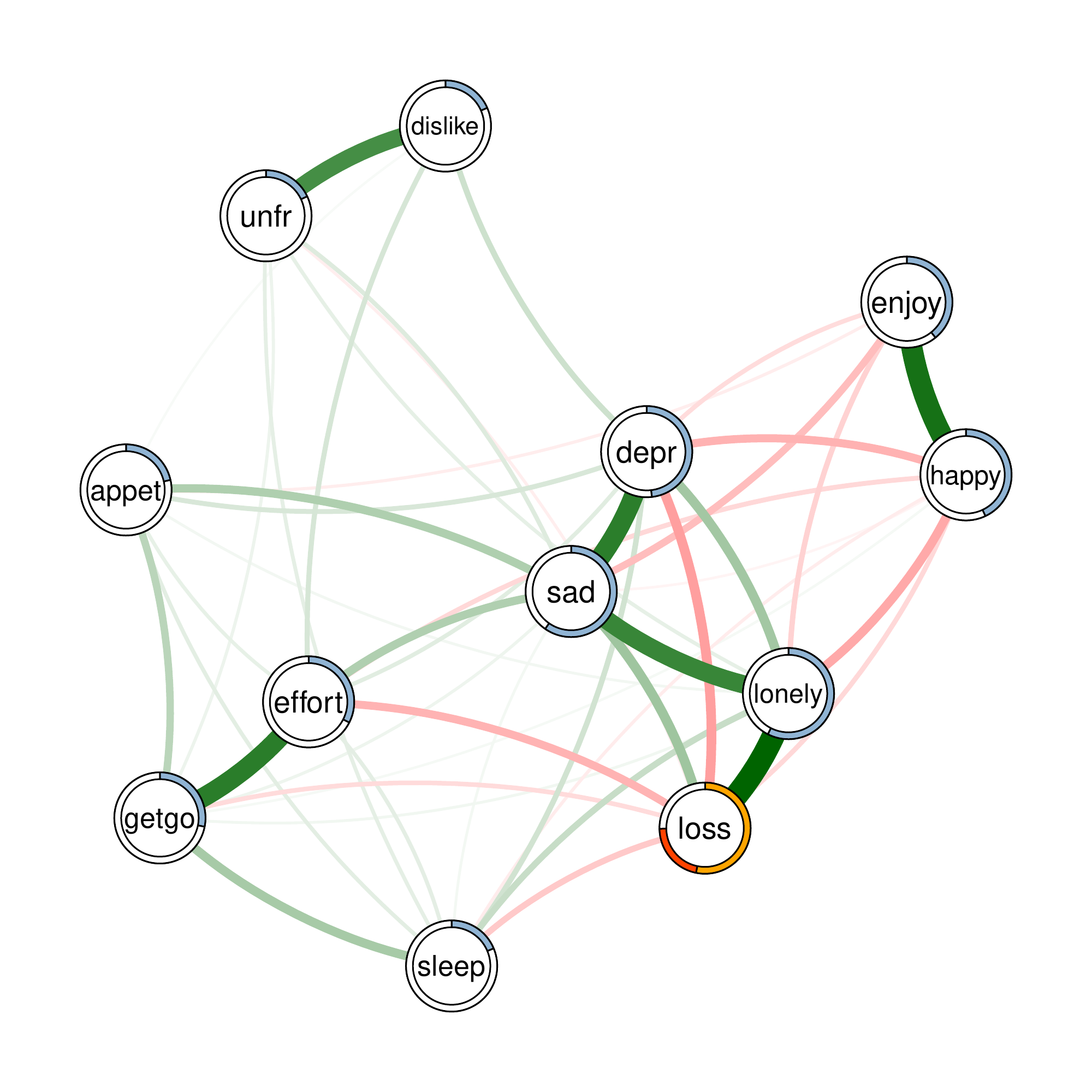}
	\caption{Mixed Graphical Model estimated on the data from Fried et al. (2015). Green edges indicate positive relationships, red edges indicate negative relationships. The blue ring shows proportion of explained variance (for continuous nodes). For the binary variable "loss", the orange part of the ring indicates the accuracy of the intercept model. The red part of the ring is the \textit{additional} accuracy achieved by all remaining variables. The sum of both is the accuracy of the full model $\mathcal{A}$. The normalized accuracy $\mathcal{A}_{norm}$ is the ratio between the additional accuracy due to the remaining variables (red) and one minus the accuracy of the intercept model (white + red).}
	\label{figure_2}       
\end{figure*}

We now provide the weighted adjacency matrix and the list containing the nodewise predictability measures to qgraph, resulting in Figure \ref{figure_2}:

\begin{Verbatim}[commandchars=\\\{\}, fontsize=\small]
pieColor <- c(rep("#90B4D4", 11), rep("#EB9446", 1)) # pick nice color

library(qgraph)
qgraph(fit_obj$pairwise$wadj, pie = error_list,
       layout="spring", labels = datalist$names,
       pieColor = color_list, abel.cex = .9, 
       edge.color = fit_obj$pairwise$edgecolor,
       curveAll = TRUE, curveDefault = .6, 
       cut = 0, labels = datalist$names)
\end{Verbatim}

The color of the pie chart behind the node can be controlled using the \verb|pieColor| argument. The remaining arguments are not necessary but improve the visualization. \verb|layout="spring"| specifies that the placement of the nodes in the visualization is determined by the force-directed Fruchtermann-Reingold algorithm \cite{fruchterman1991graph}, which places nodes such that all edges have more or less equal length and that there are as few edge crossings as possible. Note that there is no analytic relation between the distance of nodes and model parameters, however, the algorithm tends to group strongly connected nodes together in order to avoid edge crossings. Green and red edges indicate positive and negative relationships, respectively, and the width of the edges is proportional to the absolute value of the edge-weight. For a detailed description of the qgraph-package see \cite{epskamp2012qgraph}.

This code returns a network that is very similar to the one in the original paper \cite{fried2015loss}. Note that the network is not identical as we did not dichotomize ordinal variables but treat them as continuous instead. For the 11 continuous variables, the percentage of explained variance is indicated by the blue part in the pie chart. For the single binary variable, the normalized accuracy is indicated by the orange part in the pie chart.

As expected, nodes with more/stronger edges can be predicted better (e.g. \textit{lonely}) than nodes with fewer/weaker edges (e.g. unfriendly \textit{unfr}). While this trivially follows from the construction of the predictability measure (see Section \ref{section_pred_par}), this does not mean that one can use the network structure to infer the predictability of a node: by looking at the network visualization in Figure \ref{figure_2}, we  are quite certain that predictability of \textit{lonely} is higher than of \textit{unfr}. However, we do not know \textit{how} high predictability is in either of the two nodes (0.55 and 0.13, respectively), which is highly relevant for interpretation and practical applications.

Because we used the same data for estimating the network and calculating the predictability (or error) measures, we estimated the \textit{within sample prediction error}. In order to see how well the model \textit{generalizes}, one has to calculate the \textit{out of sample prediction error}. This can be done by splitting the data in two parts (or using a cross validation scheme) and providing one part to the estimation function, and the other part to the prediction function.

\section{Predictability in Temporal Networks}\label{section_temporal}

Note that the interpretation of predictability is slightly different for VAR networks because we predict each node by \textit{all} nodes at the previous time point, which also includes the predicted node itself.

\begin{figure*}
	\centering
	\includegraphics[width=0.75\textwidth]{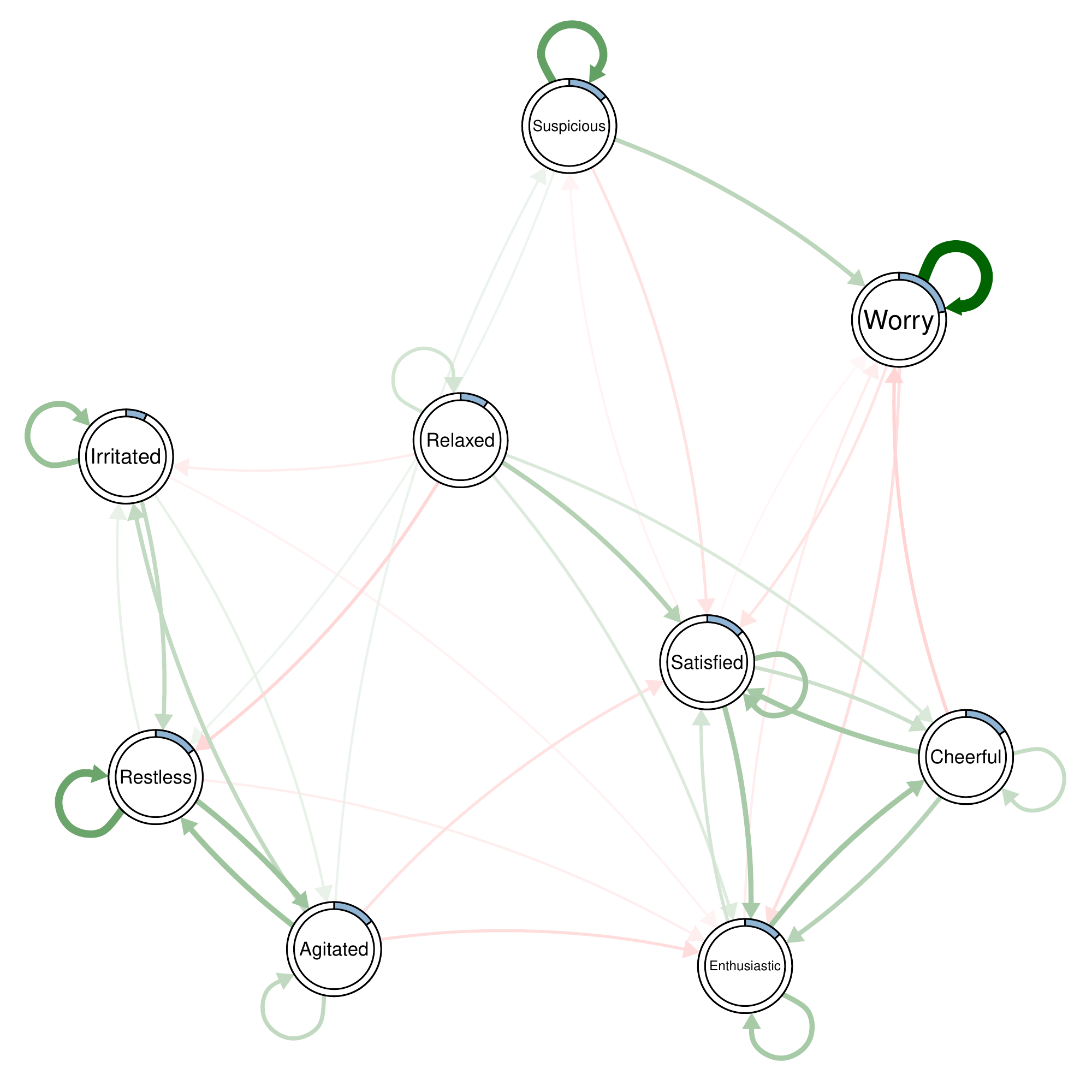}
	\caption{Visualization of VAR network of the mood variables in Wicherts et al. (2016). Green edges indicate positive relationships, red edges indicate negative relationships. The self-loops refer to the effect of the variable on itself over one time lag. The blue rings around the nodes indicate the proportion of explained variance in that node by all other nodes.}
	\label{figure_3}       
\end{figure*}

We begin again by downloading the example dataset:

\begin{Verbatim}[commandchars=\\\{\}, fontsize=\small]
url<-"https://github.com/jmbh/NetworkPrediction/raw/master/Wicherts2016_Mood.RDS"
GET(url, write_disk("Wicherts2016_Mood.RDS", overwrite=TRUE))
datalist_ts <- readRDS("Wicherts2016_Mood.RDS")
\end{Verbatim}

Next, we provide the data and the type and number of categories of variables as input. In addition, we specify that we would like to estimate a VAR model with lag 1

\begin{Verbatim}[commandchars=\\\{\}, fontsize=\small]
var_obj <- mvar(data = datalist_ts$data_mood,
		type = rep("g", 9), lev = rep(1, 9), lags = 1, 
		consec=datalist_ts$data_time$beepno)

\end{Verbatim}

\noindent
and compute the predictability of each node similarly to above:

\begin{Verbatim}[commandchars=\\\{\}, fontsize=\small]
p_obj2 <- predict(var_obj, datalist_ts$data_mood,
		  errorCon = c("R2"))
\end{Verbatim}

Finally, we visualize the network structure together with the nodewise predictability measures, which results in Figure \ref{figure_3}. Because we have only one predictability measure for each node, we can provide them in a vector via the \verb|pie| argument:

\begin{Verbatim}[commandchars=\\\{\}, fontsize=\small]
qgraph(var_obj$wadj[,,1],
       edge.color = var_obj$edgecolor[,,1],
       labels = datalist_ts$labels,
       pie = p_obj2$errors[, 2],
       pieColor = rep('#90B4D4', 9),
       curveAll = TRUE, curveDefault = .6, cut = 0)
\end{Verbatim}

We see two groups of self-engaging mood variables in Figure \ref{figure_3}: (a) the positive mood variables \textit{Cheerful}, \textit{Enthusiastic} and \textit{Satisfied} and (b) the negative mood variables \textit{Irritated}, \textit{Agitated}, \textit{Restless} and \textit{Suspicious}. \textit{Worrying} seems to be influenced by both groups and \textit{Relaxed} is rather disconnected and has a weak negative influence on group (b). These insights can be used to judge the effectiveness of possible interventions on these mood variables: for instance, if the goal is to change variables in group (a), one can do this by intervening on other variables in (a). In addition, we would expect an effect on \textit{Worrying} when intervening on groups (a) and (b), however, the reverse is not true. \textit{Relaxed} has a small influence on group (b), however, is itself not influenced by any of the variables in the network. Hence, in order to intervene on \textit{Relaxed}, one has to search for additional variables influencing \textit{Relaxed} that were not yet taken into account in the present network.

\section{Discussion}

In this paper we introduced a method and easy-to-use software to compute nodewise predictability in network models and to visualize it in a typical network visualization. Predictability is an important concept that \textit{complements} the network structure when interpreting network models: it gives a measure of how well a node can be predicted by all its neighboring nodes and is hence crucial information whenever one needs to judge the practical significance of a set of edges. An example is clinical practice, where it is important to make predictions of the outcome of interventions on an interpretable scale to optimally select treatments. 

The analyses shown in the present paper can be extended to networks that are changing over time, which allows to investigate how edge-parameters and nodewise predictability change over time. The time-varying parameters can then be modeled by a second model, which could include variables from inside and outside the time-varying network. With this modeling approach, it would be possible to gather evidence for the event of one (or several) variables causing the system to transition into another state, which is possibly reflected by a different network structure and nodewise predictability. For details about how to fit time-varying network models and time-varying predictability measures see \cite{haslbeck2015mgm}.

It is important to be clear about the limitations of interpreting nodewise predictability. First, we can only interpret the predictability of a node as the influence of its neighboring nodes if the network model is an appropriate model. A network model can be inappropriate for a number of reasons: 

\begin{enumerate}
	\item  Two or more variables in the network models are caused by a variable that is not included in the network. This results in estimated edges between these variables in the network, even though they are only related via an unobserved common cause. In this situation we cannot interpret predictability as influence by neighboring nodes, because we know that the nodes are not influencing each other but are caused by a third variable outside the network. 
	\item In some situations variables are logically dependent, for instance \textit{age} and \textit{age of diagnosis} are always related, because one cannot be diagnosed before being born. Clearly, in this situation the relation between the variables must be interpreted differently.
	\item If two or more variables measure the same underlying construct (e.g., five questions about sad mood). In this situation the edge-parameters indicate how similar the variables are and do not reflect mutual causal influence. Consequently, we would not interpret the predictability of these variables as the degree of determination by neighboring nodes. See \cite{Fried_Cramer_2016} for a discussion of this problem. Solutions could be to determine the topological overlap  \cite{zhang2005general} and choose only one variable in case of large overlap or to incorporate measurement models into the network model \cite{epskamp2016generalized}.
\end{enumerate}

Second, if we interpret the predictability of node A as a measure of how much it is determined by its neighbors, we assumed that the causal influence of the edges goes from the neighbors to node A. However, the direction of edges is generally unknown when the model is estimated from cross-sectional data. Estimates about the direction of edges can be made using causal search algorithms like the PC algorithm \cite{spirtes2000causation} or by using substantive theory. This means that the predictability of a node is an upper bound and in practice often lower, because the causal effect points away from the node at hand or is bi-directional. While this is a major limitation, note that this is true for any model estimated on cross-sectional data. In models with lagged predictors like the VAR model, this problem does not exist, because we use the direction of time to determine the causal direction.

Finally, it is important to stress that a topic we did \textit{not} cover here is to investigate how well A can be predicted \textit{by node B}.  This is different from the problem studied in this paper, where the interest was in how well node A can be predicted \textit{by all other nodes}. Unfortunately, there are no straightforward solutions for the former problem in the situation of correlated predictors, which is always the case in practice. For linear regression, there is work on decomposing explained variance \cite{gromping2012estimators} and in the machine learning literature there are methods to determine variable importance by replacing predictor variables by noise and investigate the drop in predictability \cite[e.g., ][]{breiman2001statistical}. It would certainly be interesting to try to extend these ideas to the general class of network models.

To sum up, if the network model is an appropriate model for the phenomena at hand, predictability is an easy to interpret measure of how strongly a given node is influenced by its neighbors in the network. This allows researchers to judge the practical relevance of edges connected to a node A on an absolute scale (0 = no influence on A at all, 1 = A fully determined) and thereby helps to predict intervention outcomes. In addition, the predictability of (parts of) the network is interesting on a theoretical level, as it indicates how self-determined the network is.

\begin{acknowledgements}
This research was supported by European Research Council Consolidator Grant no. 647209. JMBH would like to thank Pia Tio, Joris Broere, Max Haslbeck, Benjamin Rosche, Adela Isvoranu, Matthias Huber and Fabian Dablander for helpful comments and Sacha Epskamp for incorporating nodewise error visualizations in the qgraph package. In addition, we would like to thank two anonymous reviewers for helpful comments.
\end{acknowledgements}


\bibliographystyle{spbasic}
\bibliography{np_tech_paper}

\end{document}